\documentclass[aps, prl, amsmath, floats,floatfix, twocolumn, superscriptaddress]{revtex4-1}
 
\usepackage[normalem]{ulem}
\usepackage{amssymb}
\usepackage{amsmath}
\usepackage{verbatim}
\usepackage{mathrsfs}
\usepackage{amsfonts}
\usepackage{latexsym}
\usepackage{epsfig}
\usepackage{color}
\usepackage{graphicx,subfigure}
\usepackage{units}
\usepackage{slashbox}
\usepackage{envmath}
\usepackage{natbib}
\usepackage{hyperref}
\usepackage[utf8]{inputenc}
\usepackage{bm}

\begin{document}


\definecolor{orange}{rgb}{0.9,0.45,0}

\newcommand{\re}{\mbox{Re}}
\newcommand{\im}{\mbox{Im}}
\newcommand{\ch}[1]{\textcolor{red}{CH: #1}}
\newcommand{\er}[1]{\textcolor{cyan}{[\bf ER: #1]}}

\def\CovDev{D}
\def\Res{{\mathcal R}}
\def\Gammaflat{\hat \Gamma}
\def\metricflat{\hat \gamma}
\def\Dflat{\hat {\mathcal D}}
\def\part_n{\partial_\perp}

\def\Lie{\mathcal{L}}
\def\A{\mathcal{X}}
\def\Aphi{\A_{\phi}}
\def\hAphi{\hat{\A}_{\phi}}
\def\E{\mathcal{E}}
\def\Ham{\mathcal{H}}
\def\M{\mathcal{M}}
\def\R{\mathcal{R}}
\def\p{\partial}

\def\hg{\hat{\gamma}}
\def\hA{\hat{A}}
\def\hD{\hat{D}}
\def\hE{\hat{E}}
\def\hR{\hat{R}}
\def\hcA{\hat{\mathcal{A}}}
\def\hDelt{\hat{\triangle}}

\def\be{\begin{equation}}
\def\ee{\end{equation}}

\renewcommand{\t}{\times}

\long\def\symbolfootnote[#1]#2{\begingroup%
\def\thefootnote{\fnsymbol{footnote}}\footnote[#1]{#2}\endgroup}

\title{The non-spherical ground state of Proca stars}

 \affiliation{Departamento de Matem\'atica da Universidade de Aveiro and CIDMA,
Campus de Santiago, 3810-183 Aveiro, Portugal}

    \author{C. A. R. Herdeiro}
 \affiliation{Departamento de Matem\'atica da Universidade de Aveiro and CIDMA,
Campus de Santiago, 3810-183 Aveiro, Portugal}

  \author{E. Radu}
 \affiliation{Departamento de Matem\'atica da Universidade de Aveiro and CIDMA,
Campus de Santiago, 3810-183 Aveiro, Portugal}

\author{N. Sanchis-Gual}
\affiliation{Dept. Astronom\'{i}a y Astrof\'{i}sica, U. Val\`{e}ncia, Dr. Moliner 50, 46100, Burjassot (Val\`{e}ncia), \nolinebreak Spain}

  \author{N. M. Santos}
 \affiliation{Departamento de Matem\'atica da Universidade de Aveiro and CIDMA,
Campus de Santiago, 3810-183 Aveiro, Portugal}
 \affiliation{CENTRA, Dept. F\'\i sica, IST, Universidade de Lisboa,
Av. Rovisco Pais 1, 1049-001 Lisboa, Portugal}

  \author{E. dos Santos Costa Filho}
 \affiliation{Departamento de Matem\'atica da Universidade de Aveiro and CIDMA,
Campus de Santiago, 3810-183 Aveiro, Portugal}


\date{November 2023}

\begin{abstract}
Spherical Proca Stars (PSs) are regarded as the ground state amongst the family of PSs. In accordance, spherical PSs are thought to have a fundamental branch of stable solutions. In this \textit{Letter}, we provide energetic, morphological and dynamical evidence that spherical PSs are actually excited states. The ground state is shown to be a family of static, non-spherical, in fact \textit{prolate}, PSs. The spherical stars in the fundamental branch, albeit stable against spherical perturbations, turn out to succumb  to \textit{non-spherical} dynamics, undergoing an isometry breaking into prolate PSs. 
We also provide evidence for the dynamical formation of prolate PSs, starting from \textit{spherical} dilute initial data, via gravitational cooling. Consequently, PSs provide a remarkable example of (possibly compact) relativistic stars, in General Relativity minimally coupled to a simple, physical, field theory model, where staticity plus stability implies non-sphericity.  
\end{abstract}




\maketitle
{\bf Introduction.}
In vacuum General Relativity (GR), Israel's theorem~\cite{Israel:1967wq} establishes that \textit{staticity} implies \textit{sphericity}. This powerful result, generalizable to electrovacuum~\cite{Israel:1967za}, supports that astrophysical isolated black holes (BHs), if non-rotating, are spherical. 

Relativistic stars, on the other hand, enjoy more freedom. When matter withstands the tyranny of gravity,  richer morphologies are possible for static configurations. For concreteness, consider boson stars - self-gravitating massive, complex scalar fields in GR~\cite{Schunck:2003kk}. As static solutions, they admit a \textit{multipolar} structure~\cite{Herdeiro:2020kvf}. In the absence of self-interactions, however, the spherical stars remain the ground state, $i.e.$ the lowest energy, dynamically most robust configuration. Thus, adding the requirement of \textit{stability} to staticity still implies sphericity. 

Boson stars, together with their vector cousin Proca stars (PSs)~\cite{Brito:2015pxa}, are widely studied models of compact objects. They are self-gravitating solitons~\cite{Kaup:1968zz}, relatable to fuzzy dark matter~\cite{Suarez:2013iw,Hui:2016ltb}, often considered as BH foils~\cite{Olivares:2018abq} and ameanable to fully non-linear numerical dynamics~\cite{Liebling:2012fv}, hence permitting a breadth of applications in  astrophysics, cosmology, strong gravity and mathematical physics. Yet, the two models are not mere copies.  PSs have distinctive properties, $e.g.$, dynamically robust spinning stars~\cite{sanchis2019nonlinear}. Studying their binary dynamics is thus justified, together with the resulting waveforms~\cite{Sanchis-Gual:2022mkk} permitting comparisons, and intriguing matchings, to real events~\cite{CalderonBustillo:2020srq}. PSs also have  
distinctive geodesic flows; unlike scalar ones, spherical PSs along the fundamental branch support accretion disks similar to those around BHs, sourcing a possible degeneracy with BH images~\cite{Herdeiro:2021lwl}. 

 In this \textit{Letter} we unveil another, unexpected, distinctive feature of PSs. We establish that PSs have a non-spherical (prolate) ground state, with the spherical stars being an excited state and decaying into the former under \textit{non-spherical} dynamics. Thus, in a remarkable turn of affairs for a strong gravity model of compact stars, for PSs, staticity plus stability implies \textit{non-sphericity}.

\begin{widetext}

\begin{figure}[h!]
\begin{center}
\includegraphics[width=0.45\textwidth]{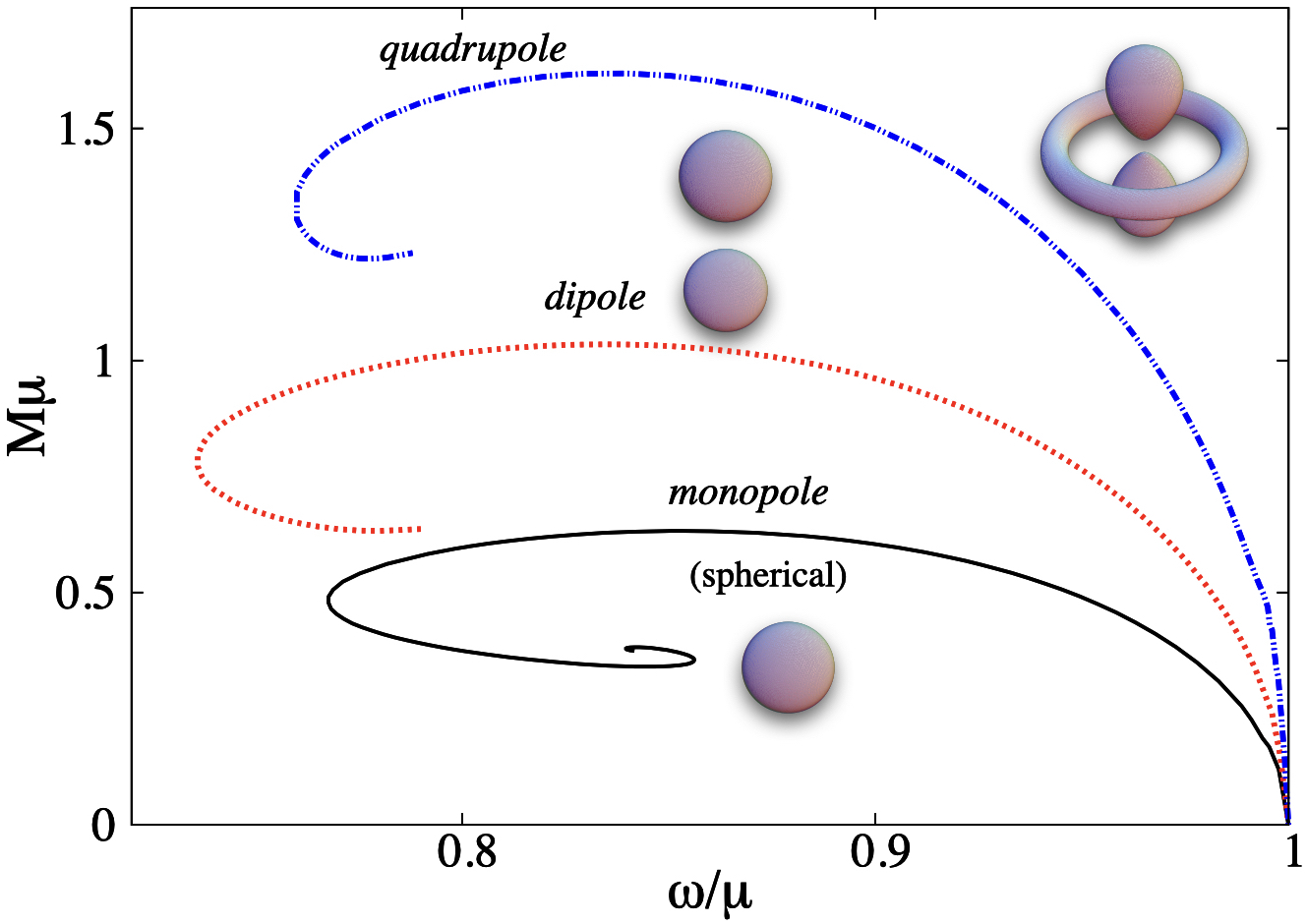} \ \ \ \ 
\includegraphics[width=0.45\textwidth]{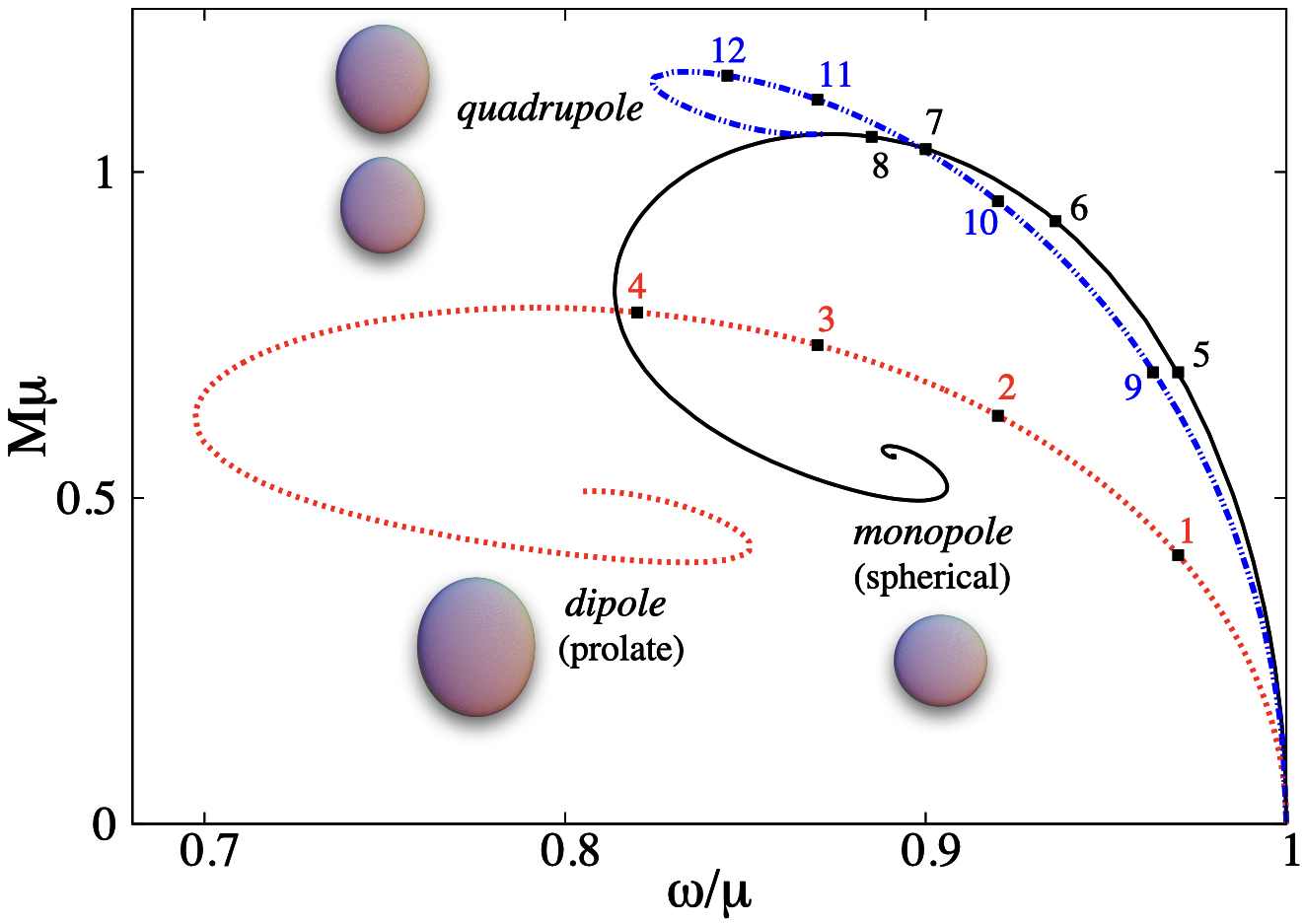}
\caption{\small Domain of existence of monopolar (spherical, $\ell=0$), dipolar ($\ell=1$) and quadrupolar ($\ell=2$) scalar boson stars (left panel) and  PSs (right panel)  in  $M$ $vs.$ $\omega$ diagrams. 
 The morphology of surfaces of constant Komar energy density is also provided. We have selected 12 illustrative solutions in the Proca case,  detailed in Table I, for the dynamical studies below.
}
\label{fig1}
\end{center}
\end{figure}

\end{widetext}

{\bf Multipolar scalar stars.}
Static scalar boson stars in the Einstein--Klein-Gordon model
    \begin{equation}
    \label{actiongen}
        \mathcal{S} = \int\text{d}^4x\sqrt{-g}\left(\frac{R}{16\pi G}+\mathcal{L}_{\rm matter}\right)\ ,
	\end{equation}
 where  $\mathcal{L}_{\rm matter}=\mathcal{L}_{\rm KG}=-\partial_\alpha \Phi \partial^\alpha \Phi^*-\mu^2\Phi\Phi^*$~\footnote{$\Phi$ denotes a complex scalar field with mass $\mu$ and complex conjugate $\Phi^*$. As usual $G$, is Newton's constant,  $g$ the metric determinant and  $R$ the Ricci scalar.}, can have a multipolar distribution - see Fig.~3 in~\cite{Herdeiro:2020kvf}.  
These multipolar stars may be seen as follows. Turning off gravity,  the Klein-Gordon equation derived from $\mathcal{L}_{\rm KG}$ on flat spacetime, in spherical coordinates, has solutions, $\Phi=e^{-i\omega t}\sum_{\ell,m}{c_{\ell m}}R_\ell(r)Y_{\ell m}(\theta,\phi)$, with radial function 
\begin{equation}
R_\ell(r)=\frac{1}{\sqrt{r}}K_{\ell+\frac{1}{2}}\left(r\sqrt{\mu^2-\omega^2}\right) \  , 
\label{flat}
\end{equation}
where $\ell \in \mathbb{N}_0$; $K_j$ are  modified Bessel functions of the second kind;  $c_{\ell m}$ are constants; the harmonic time dependence has frequency $\omega \in \mathbb{R}^+$; $Y_{\ell m}$ are real spherical harmonics. Solutions~\eqref{flat} are  irregular, diverging at the origin. But they are regularized by turning on gravity within model~\eqref{actiongen}. Their non-linear self-gravitating versions, with a single $c_{\ell m}\neq 0$, yield the multipolar stars~\footnote{These non-linear versions turn on other $Y_{\ell m}$ for a given `seed'  solution, but the original harmonic dominates asymptotically. Same holds below for the vector case.}. Focusing on the axially symmetric sector  ($m=0$),  we illustrate the domain of existence of the monopolar ($\ell=0$), dipolar ($\ell=1$) and quadrupolar ($\ell=2$) stars in Fig.~\ref{fig1} (left panel), showing also their morphology -  akin to that of hydrogen orbitals. The ADM mass $M$ increases with $\ell$, fixing $\omega$. The ground state, thus, as in hydrogen, are the spherical states $(\ell=0)$.

{\bf Multipolar PSs - flat spacetime limit.}
We now consider static PSs in the Einstein---Proca model given by~\eqref{actiongen} with $\mathcal{L}_{\rm matter}=\mathcal{L}_{\rm P}=-\frac{1}{4}\mathcal{F}_{\alpha\beta}\bar{\mathcal{F}}^{\alpha\beta}-\frac{1}{2}\mu^2\mathcal{A}_\alpha\bar{\mathcal{A}}^\alpha$~\footnote{$\mathcal{A}_\alpha$ denotes a complex Proca field with mass $\mu$, complex conjugate $\mathcal{A}_\alpha^*$ and field strenght $\mathcal{F}=d\mathcal{A}$.}. This model has a richer spectrum than its scalar counterpart. As in Maxwell's theory, it admits both electric and magnetic static solutions. Again we start by turning off gravity. The Proca equations derived from $\mathcal{L}_{\rm P}$ on flat spacetime in spherical coordinates admit the electric~\footnote{In this letter we only consider electric PSs; the magnetic case will be considered in a follow up paper.}, axiallly symmetric solutions described by the Proca ansatz
\begin{equation}
    \label{procapot}
    \mathcal{A}=e^{-i\omega t}\left(iH_0(r,\theta)\text{d}t+\frac{H_1(r,\theta)}{r}\text{d}r+H_2(r,\theta)\text{d}\theta\right)\ ,
\end{equation}
where $H_i(r,\theta)=\sum_\ell c_{i,\ell} h_{i,\ell} (r)Y_{\ell 0}(\theta)$, for $i=0,1$ and $H_2(r,\theta)=\sum_\ell c_{2,\ell} h_{2,\ell}(r)dY_{\ell 0}(\theta)/d\theta$. The radial function $h_{0,\ell}(r)=R_\ell(r)$ is given  by~\eqref{flat}~\footnote{The expressions for $h_{i,\ell}(r)$, $i=1,2$ are more involved but can also be obtained analytically.};  $c_{i,\ell}$, $i=0,1,2$ are constants. 
Again, these solutions are  irregular at the origin; and again they are regularized by turning on gravity within model~\eqref{actiongen}. Their non-linear self-gravitating versions, with ${c}_{i,\ell}\neq 0$ for a single $\ell\in \mathbb{N}_0$, yield multipolar, axially symmetric PSs.

{\bf Construction of axially symmetric multipolar PSs.}
Static, axi--symmetric solutions of~\eqref{actiongen} possess two (commuting) Killing vectors, $\{\xi,\eta\}$. 
In adapted coordinates $(t,\varphi)$, $\xi=\partial_t$ and $\eta=\partial_\varphi$. It is further assumed that the solutions admit a 2--space orthogonal to $\{\xi,\eta\}$, in which mutually orthogonal, spherical--like coordinates $(r,\theta)$ are introduced so that $g_{r\theta}=0$.
A line element compatible with these assumptions reads
\begin{equation}
    \label{metric}
   \text{d}s^2=-e^{2F_0}\text{d}t^2+e^{2F_1}\left(\text{d}r^2+r^2\text{d}\theta^2\right)+e^{2F_2}r^2\sin^2\theta\text{d}\varphi^2 \ ,
\end{equation}
where $t\in(-\infty,+\infty)$, $r\in[0,+\infty)$, $\theta\in[0,\pi]$, $\varphi\in[0,2\pi)$, and $\{F_0,F_1,F_2\}$ are real functions of $(r,\theta)$. 

To solve model~\eqref{actiongen} with the ansatz~\eqref{procapot}-\eqref{metric} for asymptotically flat, axi-symmetric PSs, appropriate boundary conditions must be imposed.
In all cases here, the equatorial plane, $\theta=\pi/2$,  is a fixed plane of the $\mathbb{Z}_2$ north-south mapping $\theta\rightarrow \pi/2-\theta$. The geometry is always $\mathbb{Z}_2$-even, whereas the Proca field may be either $\mathbb{Z}_2$-even or $\mathbb{Z}_2$-odd, depending on $\ell$. 
The monopolar solutions ($\ell=0$) are well known~\cite{Brito:2015pxa}. 
The dipolar solutions ($\ell=1$) obey, at $\theta=\pi/2$,  $\partial_\theta F_i=H_0= H_1=\partial_\theta H_2= 0$, to ensure the metric (gauge potential) is $\mathbb{Z}_2$-even ($\mathbb{Z}_2$-odd). 
Quadrupolar solutions ($\ell=2$) obey, at $\theta=\pi/2$, $\partial_\theta F_i=\partial_\theta  H_0=\partial_\theta  H_1=H_2=0$ to ensure both the metric and the gauge potential are $\mathbb{Z}_2$-even. All solutions are asymptotically flat, obeying, at infinity, $r=\infty$, $F_i=H_i=0$, $i=0,1,2$. Additionally, regularity on the symmetry axis imposes $\theta=0, \pi$, $\partial_\theta F_i=\partial_\theta H_0=\partial_\theta H_1=H_2=0$, together with $F_1=F_2$ to ensure the absence of conical singularities, for both $\ell=1,2$ solutions. Finally regularity at the origin, $r=0$, requires $\partial_r F_i=H_i=0$ ($i=0,1,2$) for $\ell=1$ and $\partial_r F_i=\partial_r H_0=H_1=H_2=0$ for $\ell=2$. 

The field equations were solved numerically employing  a professional package~\cite{schoen}.
The numerical error for the solutions is estimated to be typically less than $10^{-5}$.

{\bf Energetics and morphology.}
The domain of existence of the $\ell=0,1,2$ PSs is displayed in~Fig.~\ref{fig1} (right panel).
For all $\ell$: $(i)$ the bound state condition $\omega<\mu$ is satisfied; $(ii)$ as $\omega\rightarrow\mu$ (Newtonian limit), $M$ vanishes; $(iii)$ $M$ increases monotonically as $\omega$ decreases, reaching a maximum $M_{\rm max}$ at frequency  $\omega(M_{\rm max})$ - this interval shall be dubbed the \textit{fundamental branch}, and it contains stable solutions for ground state models; $(iv)$ further decreasing $\omega$, the mass decreases until a backbending is reached at $\omega_{\rm min}$. For $\ell=0,1,2$, in units of $\mu$, $\{M_{\rm max},\omega(M_{\rm max}),\omega_{\rm min}\}=\{1.06,0.875,0.814\}$, $\{0.792,0.792,0.698\}$ and  $\{1.15 ,0.833 ,0.824\}$.

Comparing the different $\ell$ curves  in Fig.~\ref{fig1}, within each model, unveils a fundamental difference between the Proca and scalar cases. For the former, the lowest energy configuration for a fixed  frequency within the fundamental branch is not the $\ell=0$, but rather  $\ell=1$ PSs. In fact, even $\ell=2$ stars have lower energies than $\ell=0$, except in a vicinity of $\omega(M_{\rm max})$ for $\ell=0$. This is \textit{energetic} evidence that prolate PSs are the family's ground state. 

\begin{figure}[h!]
\begin{center}
\includegraphics[width=0.45\textwidth]{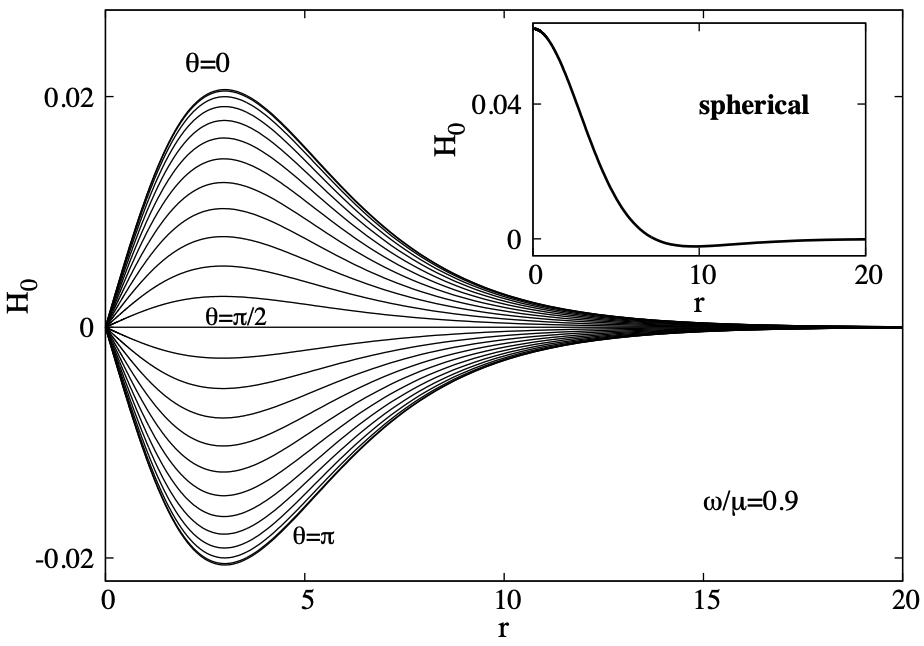}
\caption{\small  Prolate/spherical (main panel/inset) PSs do not have/have nodes. Both illustrative solutions have $\omega/\mu=0.9$.
}
\label{fig2}
\end{center}
\end{figure}

The discovery of the prolate stars also solves a  conundrum of the static PS model. On the one hand, for spherical scalar boson stars, the scalar field profile  has no radial nodes  for the ground state. Spherical excited stars are obtained by increasing the number of nodes, and have a larger energy for the same frequency - see $e.g.$ Fig.~1 in~\cite{Brito:2023fwr}. This is akin to $Ns$ orbitals in hydrogen, $N\in \mathbb{N}$. For spherical PSs, on the other hand, it can be proven that the Proca scalar potential~\cite{di2018dynamical} has no nodeless solutions~\cite{Brito:2015pxa}. Consequently, the ground state of static PSs has been regarded as the 1-node spherical solutions. But for stationary, rotating PSs there are nodeless solutions~\cite{Herdeiro:2017phl,Herdeiro:2019mbz}, which are actually dynamically robust~\cite{sanchis2019nonlinear}. Hence the conundrum: what prevents nodeless static PSs? The answer turns out to be \textit{sphericity}, since the prolate PSs have no nodes - Fig.~\ref{fig2}; they are the missing link. This is \textit{morphological} evidence they are the ground states of the static PSs family.

\begin{figure}[h!]
\begin{center}
\includegraphics[width=0.45\textwidth]{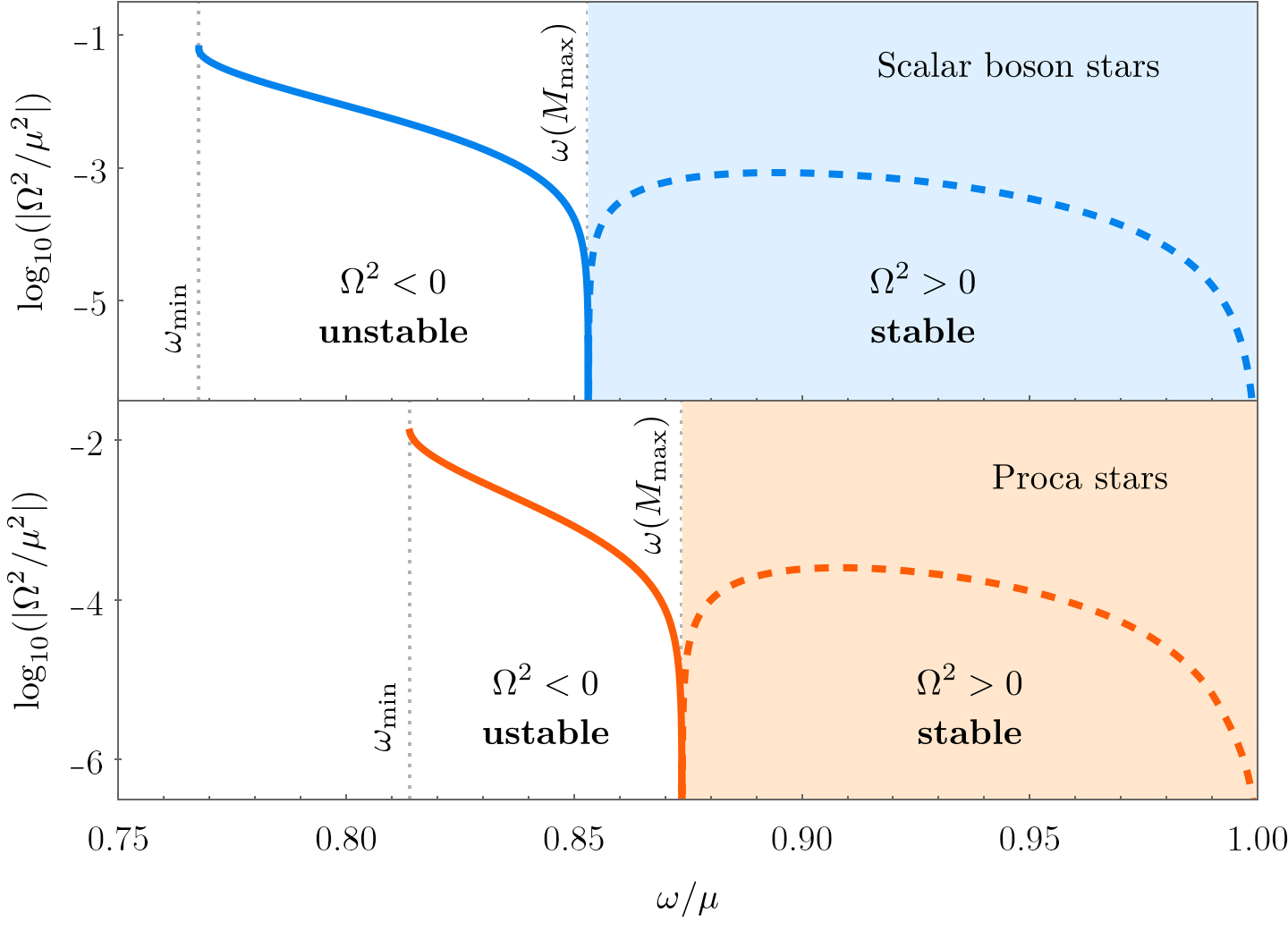}
\caption{\small Frequency squared $\Omega^2$ for spherical perturbations of monopolar ($\ell=0$) scalar (top) and Proca (bottom) stars. Dashed (solid) lines are stable (unstable) stars.
}
\label{fig3}
\end{center}
\end{figure}

{\bf Dynamics.}
Both the scalar boson stars (with no nodes) and the PSs (with one node)  in the fundamental branch are stable against \textit{spherical} perturbations. Considering a perturbation of frequency $\Omega$~\cite{Gleiser:1988ih,Hawley:2000dt,Brito:2015pxa}, linear stability changes when $\Omega^2$ changes sign~\footnote{In Fig.~\ref{fig3} we have extended the results for perturbations of PSs away from the maximal mass, the only ones reported in~\cite{Brito:2015pxa}.}. In Fig.~\ref{fig3}, $\Omega^2$ is plotted against $\omega$ for the set of scalar and PSs between the maximum and the minimum frequency. It shows that $\Omega^2>0$ for  the set of stars in the fundamental branch (dashed lines) for \textit{both} scalar and PSs. Thus, these spherical stars \textit{appear} to be the ground state for \textit{both} models. But prolate stars have lower energy in the Proca family. Thus, non-spherical dynamics has to be analysed.

Performing 3D numerical relavitity evolutions without imposing any spatial symmetries (unless otherwise stated), using the $\ell=0,1,2$ PSs as initial data, allows tackling  both the issue of stability and the end point of unstable stars. These simulations use the \textsc{einstein toolkit}~\cite{EinsteinToolkit:web}, together with the formalism and infrastructure developed in previous works~\cite{ZilhaoWitekCanudaRepository,Canuda_2020_3565475,sanchis2019nonlinear,Zilhao:2015tya}. Twelve illustrative PS solutions were selected, highlighted in Fig.~\ref{fig1} (right panel). Their mass, frequency and dynamical fate are summarized in Table I.

\begin{table}[h!]
\begin{center}
\begin{tabular}{ || c || c || c ||   }
\hline
 Dipolar/{\bf P}rolate &  Monopolar/{\bf S}pherical & {\bf Q}uadrupolar\\
\hline
\hline
{\bf 1}-(0.970,0.412) & {\bf 5}-(0.970,0.693) & {\bf 9}-(0.965,0.693) \\
\textit{Stable} & \textit{S$\rightarrow$P}   &  \textit{Q$\rightarrow$P}  \\
\hline
{\bf 2}-(0.920,0.626) & {\bf 6}-(0.936,0.924) & {\bf 10}-(0.920,0.955) \\
\textit{ Stable} & \textit{S$\rightarrow$P}  ; $[\mathbb{Z}_2]$ {\textit{S$\rightarrow$P  $(xy)$}} &  \textit{Q$\rightarrow$P}   \\
\hline
{\bf 3}-(0.870,0.735) & {\bf 7}-(0.900,1.04)& {\bf 11}-(0.870,1.11) \\  
\textit{Stable} & \textit{S$\rightarrow$P} ; $[\mathbb{Z}_2]$ {\textit{S$\rightarrow$P  $(xy)$}} &  \textit{Q$\rightarrow$BH}   \\
\hline
{\bf 4}-(0.820, 0.785) & {\bf 8}-(0.885,1.05) & {\bf 12}-(0.845,1.15)\\
\textit{Stable} & \textit{S$\rightarrow$P}   &  \textit{Q$\rightarrow$BH}  \\
\hline
\end{tabular}
\caption{\label{demo-table} Models \#-$(M,\omega)$, with \#$=1-12$ and their evolution. $[\mathbb{Z}_2]$ indicates a north-south symmetry is imposed.}
\end{center}
\end{table}

\begin{widetext}

\begin{figure}[h!]
\begin{center}
\includegraphics[width=0.9\textwidth]{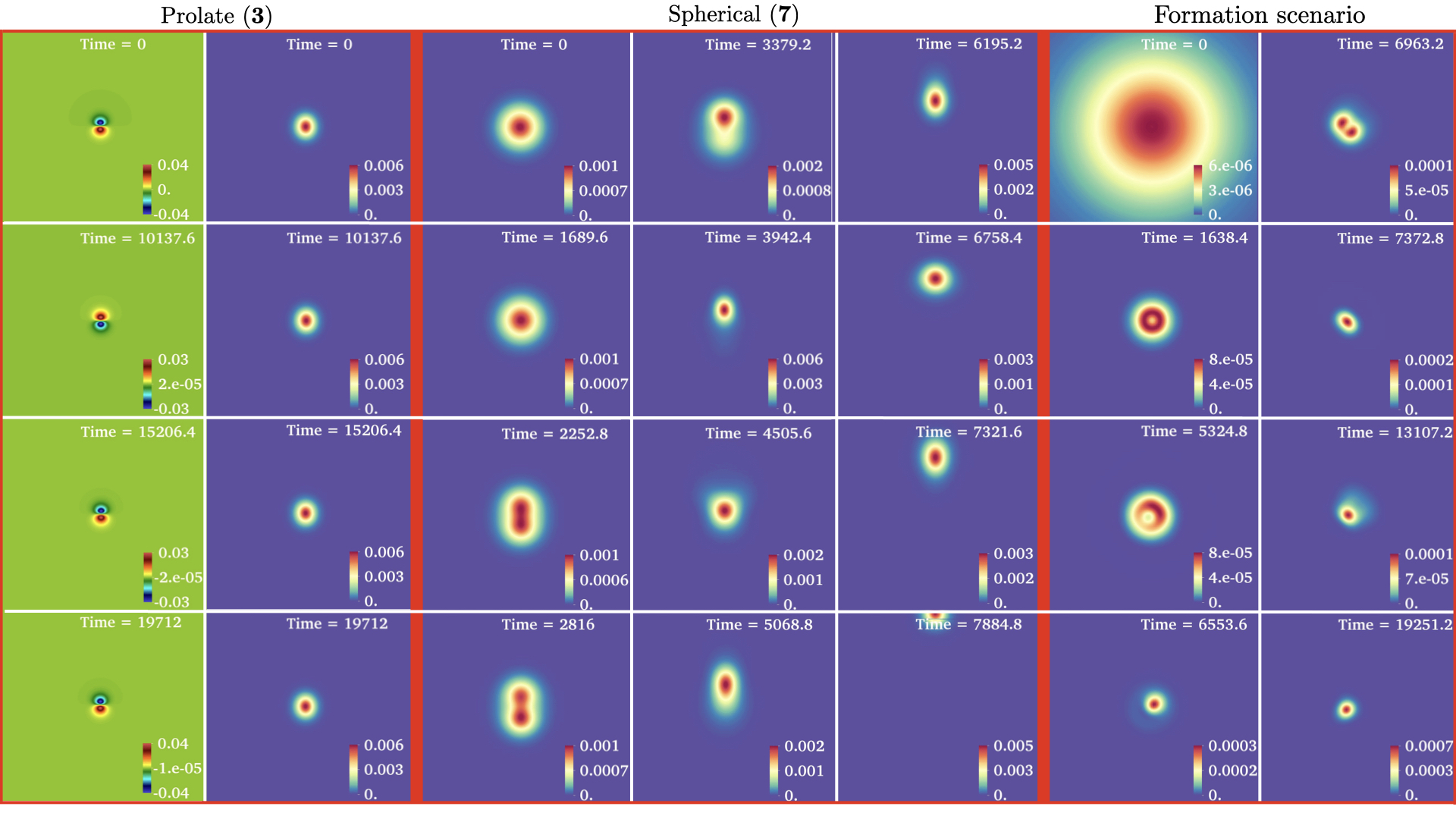}
\caption{\small Snapshots in the $x$-$z$ plane of   the imaginary part of the Proca scalar potential~\cite{di2018dynamical} (green panels) and energy density (blue panels) in PS evolutions. (Two leftmost columns) Solution {\bf 3}, illustrating the stability of prolate PSs. (Three central columns)  Solution {\bf 7}, illustrating the isometry breaking of spherical PSs into prolate stars, accompanied by a kick. (Three rightmost columns) Formation scenario, illustrating the formation of a prolate star from an initial dilute spherical Proca cloud. 
}
\label{fig4}
\end{center}
\end{figure}

\end{widetext}

We first analysed if prolate PSs are  stable in long term numerical evolutions. All prolate PSs evolved ({\bf 1}-{\bf 4} in Fig.~\ref{fig1}) have exhibited stability throughout the time evolutions performed (up to $t=2\times 10^4$, in units of $\mu$) - Fig.~\ref{fig4} (left columns). This corroborates that prolate PSs in the fundamental branch are stable.

Second, we assessed the stability of spherical stars. The evolution of solutions {\bf 5}-{\bf 8} shows that at timescales of $t\sim$ 2000 the spherical stars develop a 2-centre morphology along the $z$-axis; one of these centres swiftly dissipates leaving an oscillating prolate star, which is kicked in the $z$-direction - Fig.~\ref{fig4} (middle columns). The prolateness of the kicked star is confirmed by observing the scalar potential acquires a dipolar configuration as in the leftmost column of  Fig.~\ref{fig4}.  This corroborates the spherical PSs are decaying into the $\ell=1$ prolate stars.

The aformentioned kick is generic in the $\ell=0\rightarrow \ell=1$ decay.  Both $\ell=0$ and $\ell=1$ PSs have well defined Proca field parity, even or odd, respectively.  Thus, its energy-momentum tensor  (roughly  the Proca field squared) has even parity, sourcing a north-south $\mathbb{Z}_2$ symmetric metric.

\noindent  The instability of spherical PSs means, however, a dipolar mode grows around a spherical PS. The resulting superposition has  both even and odd parity terms, thus sourcing a non-$\mathbb{Z}_2$ symmetric geometry. The transition from a $\mathbb{Z}_2$ to a non-$\mathbb{Z}_2$ symmetric geometry generates a kick of the remnant (proto-)prolate PS~\footnote{At the linear level, this kick can be interpreted as a translation of the system's centre of mass - see Appendix G in~\cite{Zerilli:1970wzz}.}.

We have also imposed an artificial $\mathbb{Z}_2$ symmetry in some evolutions ($e.g.$ of  {\bf 6} and  {\bf 7}) effectively evolving a single hemisphere, to attempt blocking the  $\ell=0\rightarrow \ell=1$ decay. A cunning dynamical adaptation ensued.  As before, the spherical stars develop a 2-centre morphology along the $z$-axis; these 2-centres oscillate, merging and separating along the $z$ axis; but eventually the merger is followed by a  separation along the \textit{$x$-axis}. A similar dynamics to that in Fig.~\ref{fig4} (middle panels) follows, but now in the $xy$ plane, to comply with the artificial symmetry imposed, and \textit{still} generating a prolate PS. 

Concerning the $\ell=2$ PSs, we find that the evolutions of solutions {\bf 9} and  {\bf 10} form $\ell=1$ PSs, roughly resembling the evolution of the spherical PS in Fig.~\ref{fig4} starting from the (2-centre) 3$^{\rm rd}$ panel. The higher energy solutions solutions {\bf 11} and  {\bf 12}, by contrast, collapse into BHs. 

Finally, we considered the \textit{formation} of a prolate PS from a dilute \textit{spherical} distribution of Proca field. Such simulations were previously performed assuming spherical symmetry~\cite{Seidel:1993zk,di2018dynamical}, providing evidence that spherical scalar and PSs form via gravitational cooling. Following closely~\cite{di2018dynamical} but imposing \textit{no} symmetries, one observes an isometry breaking in the cloud collapse  - Fig.~\ref{fig4} (right columns) - supporting the formation of a prolate PS.

{\bf Remarks.}
Static PSs include multipolar solutions, exemplified by the $\ell=1,2$ PSs, reported here. A much larger spectrum of solutions exists, both electric and magnetic, and also non-axisymmetric, deserving detailed construction and analysis. Quite unexpectedly, and differently from the scalar model, the energetic, morphological and dynamical evidence presented herein establishes that the PSs family has a non-spherical ground state in the static sector. The $\ell=0$ PSs, considered hitherto as the ground state, succumb to non-spherical dynamics.
Spherical PSs attain their maximum energy density  away from the origin, unlike scalar stars in the ground state - see Fig.~3 in~\cite{Herdeiro:2017fhv}. For prolate PSs, the energy density is attained at the centre, befitting typical ground states. Still, a deeper understanding of why PSs dynamically prefer less isometric lumps would be desirable. One remark in this direction is that the symmetry loss in the spherical$\rightarrow$prolate decay seen here is reminscent of the fate of charged liquid drops in fluid dynamics: beyond the Rayleigh limit~\cite{doi:10.1112/plms/s1-11.1.57,doi:10.1080/14786448208628425} spherical drops decay into non-spherical ones, within a multipolar spectrum~\cite{doi:10.1063/1.857551}, as shown by experiment~\cite{nature} and simulations~\cite{PhysRevLett.106.144501}.

\newpage

{\bf Acknowledgements.}
We would like to thank Richard Brito for discussions on the perturbation theory of PSs. This work is supported by the Center for
Research and Development in Mathematics and Applications (CIDMA) through the Portuguese Foundation for Science and Technology (FCT – Fundação para a Ciência e a Tecnologia), references UIDB/04106/2020 and UIDP/04106/2020. The authors acknowledge support from the projects PTDC/FIS-AST/3041/2020, as well as CERN/FIS-PAR/0024/2021 and 2022.04560.PTDC. This work has further been supported by the European Union’s Horizon 2020 research and innovation (RISE) programme H2020-MSCA-RISE-2017 Grant No. FunFiCO-777740 and by the European Horizon Europe staff exchange (SE) programme HORIZON-MSCA-2021-SE-01 Grant No. NewFunFiCO-101086251. NSG is supported by the Spanish Ministerio de Universidades, through a Mar\'ia Zambrano grant (ZA21-031) with reference UP2021-044, funded within the European Union-Next Generation EU. This work is also supported by the Spanish Agencia Estatal de Investigacion (Grant PID2021-125485NB-C21) funded by MCIN/AEI/10.13039/501100011033 and ERDF A way of making Europe. NSG thankfully acknowledges the computer resources at Tirant and the technical support provided by UV (FI-2023-2-0002). N. M. S. is supported by the FCT
grant SFRH/BD/143407/2019. E.S.C.F. is supported by the FCT grant PRT/BD/153349/2021 under the IDPASC Doctoral Program. Computations were performed at the ARGUS cluster at Aveiro University.

\bibliography{num-rel2}


 
\end{document}